\documentstyle[prl,aps]{revtex}
\begin{document}
\draft\onecolumn
\baselineskip=12pt

\title{\bf  Optomechanical Cooling of a Macroscopic Oscillator 
by Homodyne Feedback }

\author{Stefano Mancini, David Vitali and Paolo Tombesi}

\address{Dipartimento di Matematica e Fisica, 
Universit\`a di Camerino, 
I-62032 Italy\\
and Istituto Nazionale di Fisica della Materia, Italy}

\date{Received: \today}

\maketitle
\widetext

\begin{abstract}
We propose a simple optomechanical model in which a mechanical 
oscillator quadrature could be 
"cooled" well below its equilibrium temperature by applying a 
suitable 
feedback to drive the orthogonal quadrature by means of the 
homodyne current of
the radiation
field used to probe its position.
\end{abstract}

\pacs{PACS numbers(s): 03.65.Bz, 42.50.Dv, 42.50.Vk}

\narrowtext
The problem of considering a macroscopic oscillator in terms of
Quantum Mechanics is  usually  avoided because one can obtain
the right results without using 
any quantum mechanical hypothesys. When, however, one whishes to
use it as a device to detect extremely small displacements 
due to very
weak forces, as in the gravitational wave detectors, one has 
to be careful 
in considering it as a mere macroscopic object.
Should one consider a macroscopic oscillator as a 
quantum oscillator,
once all other possible noise sources were eliminated by using
filters, screens, insulators etc.,
the ultimate criterion one has to
satisfy is the one associated with the
thermal noise \cite{Brag,thor}. For the harmonic oscillator 
it means 
$k_BT<\hbar\omega_m/2$, where $k_B$ is the
Boltzmann constant, $\omega_m$ the mechanical angular 
frequency and $T$ the
temperature of the environment in which the oscillator lives. 
This prohibitive limit, for macroscopic massive oscillators,
is however only
valid when the measurement time
$\tau$ is of the order of the mechanical relaxation time $\tau_m$.
The actual limit can be expressed as 
$2k_BT\tau/Q_m<\hbar$ \cite{Brag}. In this case it is possible 
to consider a macroscopic
mechanical oscillator as a quantum oscillator even at liquid 
He temperature \cite{BM}, 
but very high mechanical $Q_m=\omega_m\tau_m$ factors and also 
short observation times should be considered. 
To have better results and, for instance, to detect 
millisecond duration bursts of
gravitational waves from supernovae, one should 
 measure out 
of resonance, as in VIRGO or LIGO proposals \cite{virgoligo},
or at lower temperatures, as in massive 
bar detector schemes 
\cite{bars}. The thermal fluctuations are, however, the fundamental
limitations
and, in order to reduce their effects, one usually 
should lower the 
environment's temperature.

In this letter we present an alternative 
way of cooling the oscillator's
observed quadrature which could be experimentally accessible. 

We consider an empty Fabry-Perot  cavity with one 
fixed mirror 
with transmittivity $T_r$ and one perfectly reflecting 
end mirror.  
The completely reflecting
mirror can move, undergoing harmonic oscillations damped 
by the 
coupling to a thermal bath in
equilibrium at temperature $T$. The cavity resonances 
are calculated in the absence of
the impinging field, hence, if $L$ is the equilibrium 
cavity length, 
the resonant frequency of the cavity
will be
$\nu_c=\omega_c/2\pi=nc/2L$,
where $n$ is an arbitrary integer number and $c$ the 
speed of light. 
Furthermore, we assume that at the frequency of the 
impinging field  $\nu_0$, the fixed mirror
does not introduce any excess noise beyond the input field 
noise. We also assume that retardation 
effects, due to the
oscillating mirror in the intracavity field, are negligible. 
We shall use a field intensity such that
the correction to the radiation pressure force, due to the 
Doppler frequency shift of the photons
\cite{Unruh} on the moving mirror, is completely negligible. 
This means considering the damping coefficient of the 
oscillating mirror to be only due to the
coupling with the thermal bath. Thus, we can
write the Hamiltonian as
\begin{equation}\label{Hsys}
H=\hbar\omega_c(B^{\dag}B+\frac{1}{2})+\frac{{\hat p}^2}{2m}
+\frac{1}{2}m\omega_m^2{\hat x}^2+H_{int}\,,
\end{equation}
where $B$ and $B^{\dag}$ are the boson operators of the 
resonant 
cavity mode; $\hat p$ and $\hat x$ are the momentum 
and the displacement operators,
respectively, from the equilibrium position of the 
oscillating 
mirror with mass $m$ and oscillation
frequency $\nu_m=\omega_m/2\pi$.  The mechanical angular 
frequency $\omega_m$ will 
be many orders of magnitude smaller
than $\omega_c$ to ensure that the number of photons 
generated 
by the Casimir effect
\cite{Casimir} is completely negligible; we actually are 
in the so called adiabatic
approximation, i.e., the cavity round trip time of the 
photon is much shorter than the mirror's
period of oscillation.  
$H_{int}$ accounts for the interaction between the cavity 
mode and 
the oscillating mirror
\cite{Law}.  Since we have assumed no retardation effects, 
$H_{int}$ 
simply  represents the effect
of the radiation pressure force which causes the instantaneous 
displacement $\hat x$ of the mirror \cite{MT1,Pace}, and can be 
written as 
\begin{equation}\label{Hint}
H_{int}=-\hbar\frac{\omega_c}{L}B^{\dag}B\hat x=
-\hbar GB^{\dag}B(A+A^{\dag})\,,
\end{equation}
where we have introduced the dimensionless ladder 
operators ($A$ and $A^{\dag}$) 
for the oscillating mirror, 
and the coupling constant becomes
$G=\sqrt{\hbar\omega_c^2/2m\omega_mL^2}$.
The intracavity radiation field mode $B$ is damped 
through the output fixed mirror at a rate
$\gamma_b=cT_r/2L$, while $\gamma_m$ is the mechanical 
damping rate 
($\gamma_m<<\gamma_b$). 

The above interaction (\ref{Hint}) gives rise to non 
linear stochastic equations whose
linearization around the steady state is equivalent, 
in a frame rotating at the impinging frequency
$\nu_0$, to replace Eq. (\ref{Hsys}) with \cite{MT1}
\begin{equation}\label{H'}
H'=\hbar\Delta b^{\dag}b+\hbar\omega_ma^{\dag}a
+H'_{int}\,,
\end{equation}
where now all the operators represent small fluctuations 
around steady state values,
i.e. $B=\beta_s+b$ and $A=\langle A\rangle_s+a$. These
are determined by
$x_s=\sqrt{\hbar/2m\omega_m}(\langle A\rangle_s+\langle 
A\rangle_s^*)
=\hbar\omega_c|\beta_s|^2/m\omega_m^2L$;
$\beta_s=\langle B\rangle_s=\sqrt{\gamma_b}\beta_{in}/
(\frac{\gamma_b}{2}-i\Delta)$,
with $\beta_{in}$ the classical field characterizing the 
input laser power
$P_{in}=\hbar\omega_0|\beta_{in}|^2$.
The parameter $\Delta$ represents the radiation phase shift 
due to the detuning and to the
stationary displacement of the mirror,
$\Delta=\omega_c-\omega_0-\omega_cx_s/L$;
further we have 
\begin{equation}\label{hint}
H'_{int}=\hbar\chi XY_{\varphi}\,,
\end{equation}
where $X=(a+a^{\dag})/2$ is the mirror position quadrature, 
$Y_{\varphi}=
(be^{i\varphi}+b^{\dag}e^{-i\varphi})/2$ is the radiation 
quadrature with $\varphi={\rm arg}(\beta_s)$, and 
$\chi=-4G|\beta_s|$.
Eqs. (\ref{H'}) and (\ref{hint}) represent the starting 
point for further analysis of our system.

The mirror displacement induces a phase shift on 
the radiation field, hence the latter
can be used as a "meter" to measure the 
mirror position quadrature $X$.
For simplicity, in  Eq. (\ref{H'}) we assume we can 
set  $\Delta=0$ by just varying the cavity detuning.
This setting allows us to write 
the evolution 
equation for the whole density operator $D$ as
\begin{equation}
\dot{D}={\cal L}D -\frac{i}{\hbar}\left[H'_{int},D\right]
+\frac{\gamma_{b}}{2}
\left(2bDb^{\dagger}-b^{\dagger}bD-Db^{\dagger}b\right)\;,
\label{evoluzero}
\end{equation}
where ${\cal L}$ describes the damped dynamics of the 
mechanical mode $a$ 
which is considered in thermal equilibrium at temperature $T$.
We also assume that the number of
thermal photons is negligibly small at optical frequency.
Following the treatment of Refs. \cite{TV,nostro}, we
assume that the radiation mode $b$ is heavily damped, 
so that 
due to the large value of $\gamma _{b}>>|\chi|$, the $b$ mode 
always will  be 
near to its vacuum state (we are considering fluctuations 
around 
the steady state $\beta_s$). This
allows us to adiabatically eliminate the
$b$ mode  
and to perform a perturbative calculation in the small 
parameter $\chi 
/\gamma _{b}$ \cite{quadra}. 

The measurement of the quadrature $X$
is obtained by performing a homodyne measurement \cite{yuen} 
of a generic quadrature 
of the "meter" mode $b$, i.e.
$Y_{-\delta }=\left(be^{-i\delta }
+b^{\dagger}e^{i\delta }\right)/2$
with $\delta$ a phase connected to the local oscillator 
\cite{yuen}. Due to the interaction
(\ref{hint}) between the two modes, one gets information 
on $X$ by 
directly observing the mode $b$. 
The continuous monitoring of the $b$ mode via homodyne 
detection modifies 
the time evolution of the whole system.

We are now able to apply a phase-dependent feedback loop 
to control the dynamics of the mechanical mode of 
interest $a$. Other feedback schemes could be devised 
\cite{Mertz}, but 
only 
the phase-dependent feedback produces the desired effect.
In order to be easily followed, we use the continous 
feedback theory recently proposed by Wiseman and Milburn 
\cite{wisemil}, 
who well explained the implications and limitations 
of this feedback.
One has to take part 
of the stochastic output homodyne 
photocurrent \cite{Barchielli} 
obtained from the continuous monitoring 
of the meter mode $b$, and 
feed it back to the mirror dynamics 
(for example as a driving term)
in order to modify the evolution of the mode $a$. 
 
In the limiting case of a feedback
delay time much shorter than the characteristic time of the 
mechanical 
mode, it is possible to obtain a Markovian 
equation for the reduced density matrix $\rho={\rm Tr}_bD$ 
in the presence of 
feedback \cite{wisemil}. 
Thus, following Ref. \cite{wisemil}, we get
\begin{equation}
\dot{\rho }={\cal L}\rho -\frac{\Gamma }{2}\left[X,\left[X 
,\rho \right]\right]+{\cal K}\left(ie^{i\phi }\rho X-
ie^{-i\phi} X\rho\right)+\frac{{\cal K}^{2}}{2\eta\Gamma}\rho,
\label{qndfgen}
\end{equation}
where  
$\Gamma =\chi ^{2}/\gamma _{b}$ (we have defined 
$\phi =\delta +\varphi$, which is the only relevant phase of 
the $b$ 
mode influencing the dynamics of the mechanical mode $a$),
 ${\cal K}$ 
is a Liouville superoperator describing the way 
in which the feedback signal acts 
on the system of interest and $\eta$ represents 
the photodetector efficiency. 
This master equation is the starting point of our discussion.
The second term of the right hand side of Eq.~(\ref{qndfgen}) 
is the 
usual double-commutator term associated to the 
measurement of 
$X$, it results from the elimination of the radiation 
variables; the third term is the feedback term itself and 
the fourth 
term is a diffusion-like term, which is an unavoidable 
consequence of the noise introduced 
by the feedback itself.
 
The mirror is considered to be in
a thermal bath characterized by a damping constant $\gamma_m$, 
so that we have 
\begin{eqnarray}
{\cal L}\rho&=&-i\omega_m\left[a^{\dag}a,\rho\right]
+\frac{\gamma_m}{4}
\left[a+a^{\dag},\left[a^{\dag}-a,\rho\right]_+\right]\nonumber\\
&-&\frac{\gamma_m}{2}\frac{k_BT}{\hbar\omega_m}
\left[a+a^{\dag},\left[a+a^{\dag},\rho\right]\right]\,,
\label{calle}
\end{eqnarray}
where $[\;,\;]_+$ means the anticommutator and the limit 
$k_BT>>\hbar\omega_m$ 
is taken into account \cite{qnoise3}, and due 
to the frequency we are 
considering,
is surely valid at room temperature down 
to millikelvins at least.
Moreover, 
since the Liouville superoperator ${\cal K}$ can only 
be of Hamiltonian 
form \cite{wisemil}, we choose it as 
${\cal K}\rho =g 
\left[a-a^{\dagger},
\rho \right]/2$ \cite{TV,nostro},
which means feeding back the measured homodyne photocurrent to 
the mechanical oscillator with a driving 
term in the Hamiltonian involving the mechanical 
quadrature orthogonal to the
measured one; 
$g$ is the feedback gain related to the 
practical way of realizing the loop. 
One could have chosen to feed
the system with a generic phase-dependent quadrature, 
due to the homodyne current, however, it will turn
out that the above choice gives the best and simplest result.
Since the measured quadrature of the mirror is
its position the feedback will act as a drive for 
the momentum.
Using 
the above expressions in
Eq.~(\ref{qndfgen}) and rearranging the terms in 
an appropriate 
way, we finally get the following master equation:
\begin{eqnarray}\label{totale}
\dot{\rho }&=& \frac{\gamma}{2}(N+1)
\left(2a\rho a^{\dagger}-a^{\dagger}a\rho 
-\rho a^{\dagger}a\right)\\
&+&\frac{\gamma}{2}N
\left(2a^{\dagger}\rho a-aa^{\dagger}\rho 
-\rho aa^{\dagger}\right)
\nonumber \\
&-&\frac{\gamma}{2}M
\left(2a^{\dagger}\rho a^{\dagger}-a^{\dagger 2}\rho 
-\rho a^{\dagger 2}
\right)\nonumber\\
&-&\frac{\gamma}{2}M^{*}
\left(2a\rho a-a^{2}\rho -\rho a^{2}\right)
-i\omega_m\left[a^{\dag}a,
\rho \right]
\nonumber \\
&-&\left(\frac{g}{4}\sin\phi
+\frac{\gamma_m}{4}\right)
\left(\left[a^{2},\rho\right]
-\left[a^{\dag 2},\rho\right]\right)\,,\nonumber
\end{eqnarray}
where $\gamma =\gamma_m -g\sin\phi$, and
\begin{eqnarray}\label{parameters}
N&=&\frac{1}{\gamma }\left[\gamma_m\left(
\frac{k_BT}{\hbar\omega_m}
-\frac{1}{2}\right)+\frac{\Gamma 
}{4}+\frac{g^{2}}{4\eta\Gamma}+\frac{g}{2}
\sin\phi\right]\,\nonumber\\
M&=&-\frac{1}{\gamma }\left[\gamma_m
\frac{k_BT}{\hbar\omega_m}+
\frac{\Gamma}{4}-\frac{g^{2}}{4\eta\Gamma}
-i\frac{g}{2}\cos\phi\right]\,\nonumber.
\end{eqnarray}

This Eq.~(\ref{totale}) is very instructive because 
it clearly shows the 
effects of the feedback loop on the 
mirror mode $a$. 
The proposed
feedback mechanism, indeed, 
not only introduces a driving term to the mirror's momentum 
quadrature, it also
simulates the presence of a bath with nonstandard 
fluctuations, characterized by an effective damping 
constant $\gamma $ and by the coefficients
 $M$ and $N$, which are given 
in terms 
of the feedback parameters 
\cite{nostro}. For the positivity of the density matrix the
external parameters should be chosen such that $|M|^2<N(N+1)$.
This can be checked with a unitary transformation 
giving Eq. (\ref{totale})
in a manifest Lindblad form \cite{lin} for the above inequality.
An interesting aspect of the effective bath described by 
the first four 
terms in the right hand side of (\ref{totale}) is that 
it is characterized by 
phase-sensitive fluctuations, depending upon the 
experimentally adjustable 
phase $\phi$. 
 
Because of its linearity, the solution of  
Eq. (\ref{totale}) can be easily obtained,
as shown in Refs. \cite{TV}, by using the normally ordered 
characteristic 
function \cite{qnoise4} and assuming
the mirror
initially in a thermal state at temperature $T$, i.e.  
$\rho (0)=\left(1-e^{-\hbar\omega_m/k_BT}\right)
\sum_n|n\rangle\langle n|
e^{-n\hbar\omega_m/k_BT}$,
where $|n \rangle$ is the number state of the mode 
$a$. 

The stationary state is reached only if the parameters $g$, 
$\phi$, $\omega_m$ and $\gamma_m$
satisfy the stability conditions
$\gamma_m-g\sin\phi>0$
and
$\omega_m^2-\gamma_mg\sin\phi>0$.
For simplicity we choose $\phi=-\pi/2$ from now on since 
this choice 
turns out to be best.
Under the stability conditions and in the long time 
limit $(t\to\infty)$
the variance of the position quadrature operator
$X=(a+a^{\dag})/2$ for the mirror is
\begin{eqnarray}\label{varX}
&&\langle X^2\rangle=
\frac{g^2}{8\eta\Gamma}\frac{
\gamma_m^2+\omega_m^2+\gamma_mg}{(\gamma_m+g)
(\omega_m^2+\gamma_mg)}\nonumber\\
&&+\left(\frac{k_BT}{2\hbar\omega_m}+\frac{\Gamma}{8\gamma_m}
\right)\frac{\gamma_m\omega_m^2}{(\gamma_m+g)
(\omega_m^2+\gamma_mg)}\,,
\end{eqnarray}
while for the orthogonal quadrature $P=(a-a^{\dag})/(2i)$, 
i.e. the mirror's momentum, we get
\begin{eqnarray}\label{varP}
&&\langle P^2\rangle=
\frac{g^2}{8\eta\Gamma}\frac{\omega_m^2}{(\gamma_m+g)
(\omega_m^2+\gamma_mg)}\nonumber\\
&&+\left(\frac{k_BT}{2\hbar\omega_m}+\frac{\Gamma}{8\gamma_m}
\right)\gamma_m\frac{g^2+\omega_m^2+\gamma_mg}{(\gamma_m+g)
(\omega_m^2+\gamma_mg)}\,.
\end{eqnarray}

In the case of no coupling with the cavity mode,
the above variances for the macroscopic oscillator
only consist in the thermal noise 
as one should expect. 
Whenever an indirect detection of the mirror position is made 
the backaction noise is added.
The latter, however, is usually negligibly 
small compared with  the previous one.
Instead, by using the feedback with sufficient 
high gain ($g>>\omega_mQ_m$), 
we can set 
$T_{eff}\approx T\omega_m^2/g^2$ as an effective temperature,
and the mirror's position quadrature 
variance becomes
\begin{equation}\label{varghigh}
\langle X^2\rangle\approx
\frac{k_BT_{eff}}{2\hbar\omega_m}
+\frac{\Gamma\omega_m^2}{8\gamma_m g^2}
+\frac{g}{8\eta\Gamma}\,.
\end{equation}
Although the feedback 
introduces excess noise,
it also 
gives a scale factor for the thermal noise term
by means of $T_{eff}$, so that with an appropriate 
choice of the parameters, the latter can be strongly reduced.

It is also to remark that the proposed 
phase-dependent feedback
does not produce a proper squeezing; moreover, it can
extract the thermal noise from the system, 
because the variance 
reduction occurs, for not extremely high values of $g$, 
in both quadratures, as can be 
evicted from Eqs. (\ref{varX}, \ref{varP}). Hence,
it acts as a refrigerator.

To better show the potentiality of this feedback 
mechanism let us 
consider the spectrum of the position quadrature. 
To this end the Fourier 
transforms of the stochastic equations connected 
with the master equation 
(\ref{totale}) are easily written down \cite{qnoise5}
\begin{eqnarray}\label{fouriereqs}
i\omega{\tilde X}(\omega)&=&\omega_m{\tilde P}(\omega)
-g{\tilde X}(\omega)
-\sqrt{\gamma}{\tilde X}_{in}(\omega)\nonumber\\
i\omega{\tilde P}(\omega)&=&-\omega_m{\tilde X}(\omega)
-\gamma_m{\tilde P}(\omega)
-\sqrt{\gamma}{\tilde P}_{in}(\omega)
\end{eqnarray}
and the 
input noise operators have the following correlations
\begin{eqnarray}\label{incorr}
\langle{\tilde X}_{in}(\omega){\tilde X}_{in}(-\omega')
\rangle
&=&\frac{1}{4}(2N+1+2{\rm Re}\{M\})\delta(\omega-\omega')         
\nonumber\\
\langle{\tilde P}_{in}(\omega){\tilde P}_{in}(-\omega')
\rangle
&=&\frac{1}{4}(2N+1-2{\rm Re}\{M\})\delta(\omega-\omega')
\nonumber\\
\langle{\tilde X}_{in}(\omega){\tilde P}_{in}(-\omega')
\rangle
&=&\frac{1}{4}(i+2{\rm Im}\{M\})\delta(\omega-\omega')\,.
\end{eqnarray}
Defining $S_g(\omega)=
\langle{\tilde X}(\omega){\tilde X}(-\omega)
\rangle_s$, we get from Eqs. (\ref{fouriereqs})
\begin{eqnarray}\label{spectrum}
S_g(\omega)&=&
\frac{\gamma}{4}\frac{1}{|\Xi(\omega)|^2}\Bigg[
(\gamma_m^2+\omega^2+\omega_m^2)(2N+1)\nonumber\\
&+&(\gamma_m^2+\omega^2-\omega_m^2)2{\rm Re}\{M\}
\Bigg]\,,
\end{eqnarray}
where the subscript $s$ indicates the symmetrized 
correlation and 
$\Xi(\omega)=\left[(i\omega+g)(i\omega+\gamma_m)
+\omega_m^2\right]$.

As a practical example we take the
physical parameters of the model presented in
Ref. \cite{Pace}. Taking their values for granted, i.e.
$m=10$ Kg, $\nu_m=10$ Hz, $\gamma_m=1$ ${\rm s}^{-1}$, 
$L=4$ m,
$\nu_0=5.82\times 10^{14}$ Hz,
$T_r=0.02$, $P_{in}=10$ W,
we get $\Gamma\approx 200$ ${\rm s}^{-1}$, then 
$\chi\approx 10^4$ ${\rm s}^{-1}$. 
This choice satisfies the relation
$\gamma_b>>\chi$ and all  others inequalities, 
and  we further take  
$\eta\approx 1$ and $T=300\;K$.

Then, in Fig. 1 we show the (scaled) spectrum of 
Eq. (\ref{spectrum})
for various values of $g$. 
The curve for $g=0$ practically coincides with the 
spectrum of the mirror not coupled to the cavity
mode because of the smallness of the backaction noise. 
It results evident that for high values of 
the feedback gain the 
spectrum is practically
vanishing while the peak at the mechanical resonance 
frequency gradually disappears and one
peak at zero frequency appears, with very small amplitude.
With the proposed feedback, a transition
from a dissipative to a diffusive behaviour of 
the oscillator is
obtained by just varying the feedback gain.
These results, although in a different context,  
are similar, but not equivalent, to those obtained 
in Ref. \cite{Mertz},
where the feedback was used for the regulation of  
a microcantilever response and a direct 
photodetection was used,
instead of our phase-dependent scheme. Furthermore, the
temperature of the bath was assumed negligibly small.

Summarizing, we have proposed a feedback scheme based 
on an indirect measurement to reduce position
quadrature uncertainty of a macroscopic oscillator.
The described mechanism can be very useful
in reducing the effect of thermal noise in quadratures of
 macroscopic mirrors,  as those devoted to
the gravitational wave detectors, even at room temperatures.
The feedback loop may consist in a transducer 
\cite{thor} which 
transforms the random optical signal in a stochastic
 electric signal which in turn acts as a mechanical driving 
on the mirror's momentum. This could be readly realized, but 
it is not the only way, by using the 
feedback current to vary
 the potential of a capacitor formed by the oscillating 
mirror and a fixed plate.
On the other hand, depending on the specific 
experimental realization of 
the feedback loop there could be some limitations 
on the values of $g$.

We think that the practical 
implementation of the discussed model,
even though in a situation far from the oversimplified 
theoretical one, should be
an interesting challenge for an experimentalist, and it 
will turn out extremely useful in reducing the thermal 
fluctuations without lowering the bath temperature.

This work was partially supported by Istituto Nazionale 
di Fisica Nucleare. 
S. M. would like to thank F. Marchesoni for suggestions.
Discussions with C. Caves, C. Rapagnani, S. Schiller
and J. Mlynek are also greatly acknowledged.

\bibliographystyle{unsrt}

\newpage

FIGURE CAPTIONS

Fig. 1 The quantity 
$S_g(\omega)/(2\pi\langle X^2\rangle_{g=0})$
is plotted (in a semilog scale) versus $\omega$ 
for the following values of $g$ in ${\rm s}^{-1}$:
a) 0; b) 1; c) 10; d) $10^2$; e) $10^3$.
We have used the values of other parameters 
listed in the text.

\end{document}